\documentclass[aps,prb,singlecolumn,showpacs]{revtex4}
\usepackage{epsfig}
\usepackage{amsfonts,amsmath,amssymb}
\usepackage{graphicx}
\begin{document}
\title{\bf Introduction to polaron physics: Basic Concepts and Models}
\author{Julius Ranninger}
\affiliation{Centre de Recherches sur les Tr\`es Basses
Temp\'eratures, Laboratoire Associ\'e \`a l'Universit\'e Joseph
Fourier, Centre National de la Recherche Scientifique, BP 166,
38042, Grenoble C\'edex 9, France}
\maketitle

\section{Introduction}

Putting an electron into  a crystalline lattice, it will interact with the dynamical 
deformations of that latter. Depending on the type of material we are 
dealing with---such as  ionic polar crystals,  covalent materials or simple metals---the form of the electron-lattice coupling as well as the relevant lattice 
modes being involved in this coupling will be different. The mutual 
interaction  between the electron and the lattice deformations, which in 
general is of a highly complex dynamical nature, results in composite 
entities of electrons surrounded by clouds of virtual phonons which,  
in a dynamical way, correlate the position of the electron and the associated local 
lattice deformation. Such entities are referred to as polarons, 
which can be either spatially quite extended ``large polaron'', or rather 
constraint in space ``small polarons''. Large polarons generally are 
itinerant entities, while small polarons in real materials, have a tendency to self-trap 
themselves in form of localized states. By far the most spectacular manifestation of such 
electron-lattice interactions in the weak coupling large polaron limit is the classical 
phonon mediated superconductivity.

The earliest indications for polarons came from F centers---electrons trapped in negative ion vacancy positions in alkali halides---which 
led Landau to the concept of localized strong coupling polarons \cite{LandauPZS33}.
Subsequently, the field was developed mainly in connection with scenarios where 
the materials could be characterized by a continuous elastic medium and 
specifically for:\\
(i) ionic crystals \cite{FrohlichAP54}, where positively and negatively charged ions 
oscillate out of phase around their equilibrium positions and thus give 
rise to large electric polarization fields in form of longitudinal optical 
modes, which then strongly couple back onto the conduction band electrons,\\
(ii) covalent materials \cite{BardeenPR50}, where local dilations of the 
material couple the electrons to corresponding deformation potentials and\\
(iii) simple metals \cite{Scalapino69}, characterized by high frequency plasma modes which, 
because of strong charge density fluctuations,  renormalize into acoustic modes.

With the arrival of new materials since the late sixties, the necessity to consider the
microscopic lattice structure for the polaron formation became more and  more 
crucial and initiated the theoretical work on the transition between large and 
small polarons and the question of polaron localization.
 
After reviewing the different kinds of electron-lattice coupling leading to 
polaron formation we shall discuss the fundamental issues of the cross-over 
between large and small polarons, the question of continuous versus discontinuous 
transition and the difference of the polaron self-trapping with respect to 
localization in  systems with attractive interaction  potentials. We then present  
examples of a few decisive experiments on polarons which, early on, shed some light on 
their dynamical formation and disintegration. 

\section{The Fr\"ohlich Large Polaron}\label{FLP}

\subsection{Polarons in ionic crystals}

Let us to begin with consider the case of an ionic crystal where longitudinal 
optical phonons strongly couple to the electrons. A single electron in such a 
dielectricum induces via its charge an electric displacement field 
\begin{equation} 
{\bf D}({\bf r},{\bf r}_{\rm el})=-{\rm grad}{e\over |{\bf r} - {\bf r}_{\rm el}|}\, ,
\label{eq1}
\end{equation}
at a spatial coordinate ${\bf r}$ with ${\bf r}_{\rm el}$ denoting the position 
of the electron. This displacement field  couples to  the dynamics of the lattice which can be  described in terms of a  polarization field ${\bf P}({\bf r})$ of the dielectric medium and  which (via  Poisson's equation $div \; {\bf D}({\bf r})$~= $4\pi e\delta({\bf r} - {\bf r}_{\rm el})$) can be expressed in terms of a polarization potential $\Phi({\bf r})$ such as: 
\begin{equation}
{\bf P}({\bf r})= \frac{1}{4\pi} {\rm grad} \; \Phi({\bf r}), \quad
\Phi({\bf r_{\rm el}}) = - \frac{1}{e}\int {\rm d}^3{\bf r} \; {\bf D}({\bf r},{\bf r}_{\rm el}) \cdot 
{\bf P}({\bf r})\, .
\label{eq2}
\end{equation} 
We assume for simplicity the absence of any shear and vorticity in the elastic medium 
forming the dielectricum.

The standard way to describe the physics of polarons is to introduce 
measurable material quantities such as the frequency dependent dielectric constant $\varepsilon(\omega)$, defined by  ${\bf D} = \varepsilon(\omega) {\bf E}$, in order to  account for the dynamics of the dielectric medium via the 
frequency dependence of the polarization field, related to ${\bf E}$ via $4\pi{\bf P}({\bf r})$~= ${\bf D}({\bf r})-{\bf E}({\bf r})$. ${\bf D}({\bf r})$ being exclusively determined by the point charge is independent on $\varepsilon(\omega)$. In a crude sense one can consider the response of the dielectric medium as being given by a superposition of two contributions: (i)~a  high frequency contribution in the ultraviolet regime $\varepsilon_{\infty}$, arising from the  electron clouds oscillating around the ionic positions and (ii)~a low frequency contribution $\varepsilon_0$ in the infrared regime, arising from the oscillations of the positive and negative ions against each other.
This separates the polarization field ${\bf P}({\bf r})$ 
into  two contributions ${\bf P}({\bf r}) = {\bf P}_0({\bf r}) + 
{\bf P}_{\infty}({\bf r})$ which in the low and high frequency limits are determined by:
\begin{equation}
{\bf P}_0({\bf r}) +{\bf P}_{\infty}({\bf r})  = 
\frac{1}{4\pi}\left(1 - \frac{1}{\varepsilon_0}\right) {\bf D}({\bf r})\, ,
\label{eq3}
\end{equation}
\begin{equation}
{\bf P}_{\infty}({\bf r}) = \frac{1}{4\pi}\left(1 - 
\frac {1}{\varepsilon_{\infty}}\right){\bf D}({\bf r})\, ,
\label{eq4}
\end{equation}
considering the fact that the low frequency contribution of the polarization 
field remains unaffected 
by high frequency perturbations. The determinant contribution to the polaron 
dynamics, arising from the low frequency polarization field is hence given by 
\begin{equation}
{\bf P}_0({\bf r}) = \frac{1}{4\pi} \left( \frac{1}{\varepsilon_{\infty}}
 -\frac{1}{\varepsilon_0} \right){\bf D}({\bf r}) 
\equiv \frac{1}{4\pi\tilde{\varepsilon}}{\bf D}({\bf r})\, .
\label{eq5}
\end{equation}

Before entering into a detailed discussion of the intricate nature of 
the dynamics of this problem let us  consider certain limiting cases and 
start with the picture of an  electron inside a dielectric continuous medium 
being constraint to a finite small volume in a sphere of a certain radius $R_1$ to be determined.

We consider an electron, moving  inside such a sphere of radius $R_1$ 
with a velocity $v$, to be fast compared to the characteristic time of the atomic oscillations $(2\pi/\omega_0)$ (given by the longitudinal optical lattice modes). The polarization field induced in the 
medium by the motion of the charge carrier can then be considered as static 
for distances much greater than $2 \pi v/\omega_0$ and being described by a static Coulomb potential. Inside this sphere  the test charge can be considered as 
being uniformly distributed and hence the potential as being constant. We thus have:
\begin{equation}
E_{\rm pot} = -e^2 / \tilde\varepsilon R_1\;\; 
(r < R_1), \quad
E_{\rm pot} = -e^2 /  \tilde\varepsilon r\;\; (r \geq R_1)\, .
\label{eq6}
\end{equation}

Determining the size of the polaron in a purely static fashion, considering  the electron kinetic energy  inside this sphere as being given by 
$E_{\rm kin} \simeq (2\pi \hbar)^2/2m R^2_1)$, 
we find by  minimizing the total energy with respect to $R_1$:
\begin{equation}
R_1 = (2\pi \hbar)^2 \tilde\varepsilon / m e^2\, .
\label{eq7}
\end{equation}

Determining the size of the polaron radius in a dynamical fashion we require that the characteristic wavelength $2 \pi \hbar/mv$ of the electron must be smaller than $2 \pi v/\omega_0$. This implies that the typical radius $R_2$ of the trapping potential is determined by the distance for which $\hbar/mv \simeq v/\omega_0$, or in other words:
\begin{equation}
R_2 = 2 \pi (\hbar /m\omega_0)^{\frac{1}{2}}\, .
\label{eq8}
\end{equation}
>From these simple arguments it follows 
that the interaction between the displacement field and the polarization 
field, arising from the dynamics of the lattice, (the polar coupling between 
the electron and the longitudinal optical phonons)) is of the order of 
$E_{\rm pot}$~= $-e^2/\tilde{\varepsilon}R_2$.

The potential energies for the two limiting cases where the 
dielectric medium is considered (i)~as static and (ii)~as dynamic, are given by
\begin{equation}
-{U_1 \over \hbar \omega_0} = \left({U_2 \over \hbar \omega_0}\right)^2 =
{1 \over 2 \pi^2} \alpha^2_{\rm Fr}\, ,
\label{eq9}
\end{equation}
where $\alpha_{\rm Fr}$ is a dimensionless coupling constant introduced by Fr\"ohlich
\begin{equation}
\alpha_{\rm Fr}= {e^2 \over \tilde{\varepsilon} \sqrt2}
\sqrt{{m \over \hbar^3 \omega_0}}\, .
\label{eq10}
\end{equation}
The value for  $\alpha_{\rm Fr}$ can vary over one decade in the interval 
$\sim[0.1, 5]$ depending on the material. Large values 
of $\alpha$ favor the static picture, while smaller ones favor the 
dynamical one. For most standard materials (polar and ionic crystals) 
the typical phonon frequency is of the order of $10^{-14}$ seconds and the 
electrons can be assumed to be essentially free electrons. 
This results in a value for $R_2$ which is much bigger than the lattice 
constant and hence justifies the continuum approach described above 
permitting to treat the dielectric constant as wave vector independent.
Within such an approach Fr\"ohlich first formulated this polaron problem in a 
field theoretical form and proposed a corresponding Hamiltonian for it, 
designed to describe the so called large polarons. 

Let us now briefly sketch the derivation of the corresponding
Hamiltonian \cite{FrohlichPM50}. To within a first approximation the dynamics of the polarization field can be described by harmonic oscillators driven by the electric displacement fields. Dealing with the polaron problem on the basis of the continuum model of a dielectricum one is restricted to consider waves of the polarization field with wave length much bigger than the inter-atomic distance and moreover to limit oneself in this problem to frequencies well below the optical excitations which would result from the deformations of the ions rather than from their motion. Under those 
conditions one can limit the discussion of the dynamics of the polarization 
field to the infrared longitudinal component of it, being driven by a source term due to its coupling with the electric displacement field. Using a 
simple Ansatz in form of  a harmonic motion of the polarization field one  has  
\begin{equation}
\left({{\rm d}^2 \over {\rm d}t^2} + \omega^2_0 \right) {\bf P}_0({\bf r}) = 
{1 \over \gamma}{\bf D}({\bf r},{\bf r}_{\rm el})\, .
\label{eq11}
\end{equation}
The coefficient ${1\over \gamma}$~= $\equiv{\omega_0^2 \over 4\pi}{1\over \tilde\varepsilon}$ of ${\bf D}({\bf r},{\bf r}_{\rm el})$ on the rhs of this 
equation is determined from its static limit, following eqs.~(\ref{eq3}, \ref{eq4}). To this equation of motion for the 
polarization field one has to add the term which describes the motion of 
the electron and which is controlled by the interaction energy $e\Phi({\bf r})$, 
given in eq.~(\ref{eq2}), i.e., 
\begin{equation}
m{{\rm d}^2 \over {\rm d}t^2} {\bf r}_{\rm el} = - e \; {\rm grad} \; \Phi({\bf r}_{\rm el})\, .
\label{eq12}
\end{equation}
Considering $\gamma{\bf P}_0({\bf r})$, $m{\bf r}_{\rm el}$ and 
$\gamma\dot{\bf P}_0({\bf r})$, ${\bf p}_{\rm el}$~= $m\dot{\bf r}_{\rm el}$ as a set of generalized coordinates and conjugate momenta $\{Q,\partial L\partial \dot Q\}$ the corresponding Hamiltonian is given by
\begin{equation}
H  = \frac{1}{2}\gamma\int{\rm d}^3{\bf r}\left[\dot{\bf P}^2_0({\bf r})+\omega_0^2{\bf P}^2_0({\bf r})\right]-\int{\rm  d}^3{\bf r}{\bf D}({\bf r},{\bf r}_{\rm el}){\bf P}_0({\bf r})+\frac{1}{2}m \dot{\bf r}_{\rm el}^2\, .
\label{eq13}
\end{equation}
In order to obtain a quantum field theoretical formulation of this problem  
one introduces habitually the vector field representation for the polarization 
fields
\begin{equation}
{\bf B}^{\pm}({\bf r})=\sqrt{{\gamma \omega_0 \over 2\hbar}}\left({\bf P}_0({\bf r})\pm {i \over \omega_0}\dot{\bf P}_0({\bf r}) \right)={1 \over N} \sum_{{\bf q}\lambda}{{\bf q} \over q}\left(\begin{array}{c}
e^{-iqr}a^+_{{\bf q}\lambda}\\
e^{+iqr}a_{{\bf q}\lambda}\end{array}\right) \, ,
\label{eq14}
\end{equation}
with the phonon annihilation (creation) operators 
$a^{(+)}_{{\bf q}\lambda}$ with $[a_{{\bf q}\lambda},a^+_{{\bf q}'\lambda'}]$~= $\delta_{{\bf q},{\bf q}'}\delta_{\lambda,\lambda'}$.

For ionic polar crystals the dominant contribution of  the polarization field 
comes from longitudinal optical phonons. We shall in 
the following restrict ourselves to those modes only. Introducing the 
quantized form of the electron momentum ${\bf p}_{\rm el} \equiv {\hbar\over i}{\partial\over\partial {\bf r}_{\rm el}}$ with $[{\bf p}_{\rm el},{\bf r}_{\rm el}]$~= $-i\hbar$, we  rewrite the Hamiltonian, eq.~(\ref{eq13}) 
in terms of the phonon creation and annihilation operators  and the electron 
coordinates and conjugate momenta as
\begin{equation}
H ={p^2_{\rm el} \over 2m} + \hbar \omega_0 \sum_{\bf q} \left(a^+_{\bf q} a_{\bf q}+\frac{1}{2}\right) + \sum_{\bf q} V^{\rm opt}_{\bf q}(a^+_{\bf q}e^{-i{\bf q} \cdot {\bf r}_{\rm el}} - a_{\bf q}e^{i{\bf q} \dot{\bf r}_{\rm el}})\, ,
\label{eq15}
\end{equation} 
where the effective electron lattice coupling constant is
\begin{equation} 
V^{\rm opt}_{\bf q} = i \sqrt{4 \pi \alpha} 
(\hbar/2 \omega_0 m)^{\frac{1}{4}}\hbar \omega_0 (1/q)\, .
\label{eq16}
\end{equation}
Upon introducing dimensionless electron coordinates as well as phonon wave vectors $\bar{\bf r}_{\rm el}$~= $(2m\omega_0 / \hbar)^{\frac{1}{2}}{\bf r}_{\rm el}$  and $\bar {\bf q}$~= $(2m\omega_0 / \hbar)^{-\frac{1}{2}}{\bf q}$,  $H/\hbar\omega_0$ turns out to be exclusively controlled by the unique parameter $\alpha_{\rm Fr}$.

Replacing in the expression for the Hamiltonian, eq.~(\ref{eq15}),  the Fourier transform $e^{i{\bf q}\cdot {\bf r}}$ of the electron density 
$\rho({\bf r})$~= $\sum_{{\bf r}_{\rm el}} \delta({\bf r} -{\bf r}_{\rm el})$ by its 
second quantization form $\sum_{\bf k\sigma}c^+_{\bf k+q\sigma}c_{\bf k\sigma}$ we finally obtain what is generally called the  Fr\"ohlich Hamiltonian
\begin{equation}
H =\sum_{{\bf k}\sigma} \varepsilon_{\bf k}c^+_{{\bf k}\sigma}c_{{\bf k}\sigma}+\sum_{\bf q}\hbar \omega_0(a^+_{\bf q}a_{\bf q}+\frac{1}{2})+\sum_{{\bf q}{\bf k} \sigma}V^{\rm opt}_{\bf q}c^+_{{\bf k+q}\sigma}c_{{\bf k}\sigma}(a_{\bf q}-a^+_{-{\bf q}})\, .
\label{eq17}
\end{equation} 
$\varepsilon_{\bf k}$~= $D-t\gamma_{\bf k}$, $\gamma_{\bf k}$~= 
$\frac{1}{z}\sum_{\delta} \exp(i{\bf k} \cdot {\delta})$ denotes the bare electron dispersion (with $\delta$ being the lattice vectors linking nearest neighboring sites) and $D=zt$ is the band half-width and $z$ the coordination number.

\subsection{Polarons in non-polar covalent materials}

A structurally very similar Hamiltonian is obtained for covalent 
materials when the density of charge carriers is very low and screening
effects can be neglected. The electron lattice interaction then can be 
derived  within the so-called {\it deformation potential} method \cite{BardeenPR50} where it is assumed that the electron dispersion of the rigid lattice $\varepsilon^0_{\bf k}$ gets modified due to some elastic strain which, in the simplest case, amounts to a dilation. This then results in a shift of the electron dispersion $\varepsilon_{\bf k}$~= $\varepsilon^0_{\bf k}+C\Delta$ with
\begin{equation}
\Delta(x)={\partial u^{\nu} \over \partial x^{\nu}} = i {1 \over \sqrt N}\sum_{\bf q} {q \over \sqrt{2 M \omega_{\bf q}}} (a_{\bf q} e^{i {\bf q} \cdot {\bf x}} - 
a^+_{\bf q} e^{-i {\bf q} \cdot {\bf x}})\, .
\label{eq18}
\end{equation}
The dilation, being  exclusively related to longitudinal acoustic phonons 
($a_{\bf q}, a^+_{\bf q}$) with a phonon frequency  $\omega_{\bf q}$ and $M$ denoting the ion mass,  the constant $C$ can be estimated 
form pressure measurements and is typically of the order of $10$~eV. The 
effective electron-lattice interaction in that case is then similar to 
that derived above for ionic crystals and is described by the Hamiltonian 
eq.~\ref{eq17} upon replacing $V^{\rm opt}_{\bf q}$ by 
\begin{equation}
V^{\rm ac}_{\bf q} = i C {q \over  \sqrt{\frac{1}{2} M \omega_{\bf q}}}\, .
\label{eq19}
\end{equation}

\subsection{Polarons in metals}

A yet very different approach is required to treat the electron-lattice 
coupling in metals. The electrons at the spatial coordinate ${\bf x}$ 
experience a pseudo potential $V({\bf x}-{\bf R}_i)$ exerted on them by the 
positively charged ions situated at lattice sites ${\bf R}_i$. The major 
effect giving rise  to the coupling of the electrons to the lattice now comes from the dynamical displacements ${\bf u}_i = {\bf R}_i - {\bf R}^0_i$ of the ions from their equilibrium positions ${\bf R}^0_i$. The dynamics of the lattice arises primarily from the ion-ion interaction $\tilde{V}({\bf R}_i-{\bf R}_j)$, which can being assumed in form of a Coulomb potential. But since those ions are imbeded in a kind of electron jellium, they sense the 
charge density fluctuations of the electrons via the electron-lattice 
interaction. Considering plane waves $\psi({\bf x})$ for the electrons 
one has the following Hamiltonian to consider
\begin{eqnarray}
H &=& \int {\rm d}{\bf x} \psi^+({\bf x}) \left[ - {\nabla^2 \over 2m} + 
\sum_i V({\bf x}-{\bf R}_i)\right] \psi({\bf x}) 
+ \sum_i {P^2_i \over 2M}
+ {1 \over 2} \sum_{i,j} \tilde{V}({\bf R}_i-{\bf R}_j) \nonumber \\ 
&&+ {1 \over 2} 
\int {\rm d}{\bf x} {\rm d}{\bf x}' \psi^+({\bf x})  \psi^+({\bf x}') 
{e^2 \over |{\bf x} - {\bf x}'|}\psi({\bf x}')\psi({\bf x})\, ,
\label{eq20}
\end{eqnarray}
where $P^2_i/2M$ denotes the kinetic energy of the ions  at site $i$.
Using the standard expansion of the lattice displacements ${\bf u}_i$ 
in terms of the phonon operators 
\begin{equation} 
{\bf u}_i = {1 \over \sqrt N } \sum_{{\bf q}\lambda} {\bf e}_{{\bf q}\lambda} 
e^{i {\bf q} \cdot {\bf R}^0_i}  \left(\hbar \over 2 M 
\Omega_{{\bf q}\lambda} \right)^\frac{1}{2}(a_{{\bf q}\lambda} + 
a^+_{-{\bf q}\lambda})\, .
\label{eq21}
\end{equation}
(${\bf e}_{{\bf q}\lambda}$ denotes the polarization vectors and 
$\Omega_{{\bf q}\lambda}$ the eigen-frequencies of 
the various phonon branches) one can write the above Hamiltonian in the form
\begin{eqnarray}
H &=& \sum_{{\bf k}\sigma} \varepsilon_{\bf k} c^+_{{\bf k}\sigma} 
c_{{\bf k}\sigma} + \sum_{{\bf q}\lambda} \hbar \Omega_{{\bf q}\lambda} 
(a^+_{{\bf q}\lambda}a_{{\bf q}\lambda}+\frac{1}{2}) 
+ \sum_{\bf q}{4 \pi e^2 \over q^2}\rho_{\bf q}
\rho_{-{\bf q}}\nonumber\\
&&+ \sum_{{\bf q}\lambda}V^{\rm met}_{{\bf q}\lambda}\rho_{\bf q} 
(a_{{\bf q}\lambda} + a^+_{-{\bf q}\lambda})\, ,
\label{eq22}
\end{eqnarray}
where $\rho_{{\bf q}} = {1 \over  N}\sum_{{\bf k}\sigma} 
c^+_{{\bf k}+ {\bf q}\sigma} c_{{\bf k}\sigma}$ is the charge density fluctuation 
operator for the electrons and 
$V^{\rm met}_{{\bf q}\lambda}$~=  $-i(N/2M \hbar\Omega_{{\bf q}\lambda})^\frac{1}{2}({\bf q} \cdot {\bf e}_{{\bf q}\lambda}) V({\bf q})$.  The relevant phonon modes arising from the ion-ion interaction are the practically dispersion-less ionic plasma modes with frequencies $\Omega_{{\bf q}\lambda}$. However, these modes being coupled to the electron charge fluctuations, this results in a strong renormalization leading to: (i) a longitudinal acoustic branch with a correspondingly phonon frequency
$\omega^{ac}_{\bf q}=\simeq q \sqrt{m/3M} v_F$ ($v_F$ denoting the Fermi 
velocity) and (ii) to a corresponding dressed electron-phonon coupling 
given by \cite{Scalapino69} 
\begin{equation}
\tilde{v}_{{\bf q}\lambda} = -i \sqrt {N \over 2M\omega^{ac}_{{\bf q}\lambda}}
({\bf q} \cdot {\bf e}_{{\bf q}\lambda}) {\Lambda({\bf q})V({\bf q}) \over 
\varepsilon({\bf q})Z_c}\, .
\label{eq23}
\end{equation}
$\varepsilon({\bf q})$ denotes a wave vector dependent dielectic constant, 
$\Lambda({\bf q})$ the Coulomb screened renormalization of the bare 
ion potential and $Z_c$ the spectral weight renormalization constant of the 
electron quasi-particle  spectrum.

In spite of the diversity of the physical systems considered above, the various  Hamiltonians  describing them are  of rather general form. 
The main message which they contain is that, because of the electron-lattice 
interaction $V_{\bf q}$, the electrons will be accompanied by a 
lattice deformation. In the weak coupling limit, this is restricted to simply a 
single phonon accompanying the electron. Within a lowest order perturbative 
approach this leads to states of the form 
\begin{equation}
c^+_{{\bf k}\sigma}|0\rangle + \sum_{\bf q} V_{\bf q} c^+_{{\bf k}- {\bf q}\sigma}a^+_{\bf q}|0\rangle 
{1 \over (\hbar \omega_{\bf q} +\varepsilon_{{\bf k}- {\bf q}} - 
\varepsilon_{\bf k})}\, ,
\label{eq24}
\end{equation}
where $|0\rangle$ denotes the  vacuum state for the electrons as well as of the phonons. This expression clearly indicates that electrons  carry with them an electrical polarization field or, in other terms, the electrons are accompanied by a phonon. This feature reflects the fact that the total wave-vector 
\begin{equation} 
{\bf K} = \sum_{{\bf k}\sigma} {\bf k} c^+_{{\bf k}\sigma}c_{{\bf k}\sigma} + 
\sum_{{\bf q}\lambda}{\bf q}a^+_{{\bf q}\lambda}a_{{\bf q}\lambda}\, ,
\label{eq25}
\end{equation}
is a conserved quantity. The most spectacular consequence of this 
lies in the formation of Cooper pairs \cite{CooperPR56}, as evidenced  by the isotope effect of their binding energy \cite{FrohlichPPSA50} and the ultimately resulting phonon mediated BCS theory of superconductivity \cite{BardeenPR57}. Electron pairing then can be understood by a process in which the passage of a first electron polarizes the lattice  and where subsequently a second electron reabsorbes that polarization. The effective interaction Hamiltonian  for this pairing is given by second order perturbation theory:
\begin{equation}
H_{\rm el-el} = \sum_{\bf kk'q}|V_{\bf q}|^2
c^+_{{\bf k+q}\uparrow}c^+_{{\bf k'-q}\downarrow}c_{{\bf k'}\downarrow}
c_{{\bf k}\uparrow}{\hbar \omega_{\bf q} \over (\varepsilon_{\bf k} - 
\varepsilon_{\bf k-q})^2 -(\hbar \omega_{\bf q})^2 }\, .
\label{eq26}
\end{equation}
This effective electron-electron interaction shows that within 
a small region around the Fermi surface i.e., $|\varepsilon_{\bf k} - 
\varepsilon_{{\bf k} \pm {\bf q}}| \leq \omega_{\bf q}$ this interaction is 
attractive and therefore leads to an instability of the Fermi surface  
resulting in a superconducting ground state via Cooper pair formation in 
${\bf k}$-space.

This present discussion of the continuum approach to the coupling between 
the electrons and the lattice-deformations has shown interaction Hamiltonians which, depending on the underlying materials, show either a coupling between the phonon coordinates $(a^{\phantom +}_{\bf q} + a^+_{-{\bf q}})/\sqrt 2$ and the electron charge 
density $\rho_{\bf q}$~= $\frac{1}{N}\sum_{\bf k}c^+_{{\bf k}+ \frac{\bf q}{2}}
c^{\phantom +}_{{\bf k}+ \frac{\bf q}{2}}$ (metals) or between the conjugate phonon momenta  $(a^{\phantom +}_{\bf q} -a^+_{-{\bf q}})/i\sqrt 2$ 
and the charge current density ${\bf p}_{\bf q} = \frac{1}{N}\sum_{\bf k}{\bf q}c^+_{{\bf k}+ \frac{\bf q}{2}}c^{\phantom +}_{{\bf k}+ \frac{\bf q}{2}}$ 
(polar and covalent materials). Formally these interaction terms can be written in a unifying way by rotating the phonon coordinate  
into the phonon momenta and vice versa by a suitable unitary transformation $U =\exp(-i\frac{\pi}{2}a^+_{\bf q}a^{\phantom +}_{\bf q})$.

\section{The Holstein small polaron}\label{Hsp}

The systems discussed in the previous section are characterized
by long range electron-lattice coupling which show up in form of 
(i)~a moderate mass renormalization of charge carriers in their band 
states, (ii)~an equally moderate reduction in their mobility due to the scattering of the electrons  on the lattice vibrations and (iii) the emergence of phonon sidebands in optical absorption spectra. 

The fundamental theoretical question which posed itself in the context of 
polaron physics ever since Landau \cite{LandauPZS33} proposed self-trapped localized polarons was to establish whether in such systems one would have a local lattice instability upon increasing the coupling constant $\alpha_{\rm Fr}$, passing from large mobile Fr\"ohlich polarons for weak coupling to localized polarons when the coupling strength exceeds a certain critical value. This question could not be addressed within the continuum approach since it requires a physics which is related to 
the dynamics of the local lattice deformations on the scale of the unit 
cell. On the experimental side, more and more new materials where synthesized in the mean time which clearly showed polaronic effects on such short length scales. It was for these reasons that a scenario was introduced which could describe such lattice polarons.
 
The generic model to capture such a situation, generally referred to as the Holstein molecular crystal model \cite{HolsteinAP59} treats the problem consequently in real space. Its corresponding Hamiltonian is given by
\begin{eqnarray}
H &=& D\sum_{{\bf i}\sigma}n_{{\bf i}\sigma} - 
t\sum_{{\bf i}\neq{\bf j}\sigma}(c^+_{{\bf i}\sigma}c^{\phantom{\dag}}_{{\bf j}\sigma}+
h.c.)
-\lambda\sum_{\bf i}n_{\bf i}u_{\bf i}\nonumber \\
& & + \sum_{\bf i}\frac{M}{2}(\dot{u}^2_{\bf i}+\omega^2_0u^2_{\bf i})
+ \sum_{\bf i}Un_{{\bf i}\uparrow}n_{{\bf i}\downarrow}, \label{H}\,
\label{eq27}
\end{eqnarray}
where $n_{{\bf i}\sigma} = c^+_{{\bf i}\sigma}c^{\phantom{\dag}}_{{\bf i}\sigma}$ 
($n_{\bf i} = \sum_{\sigma}n_{{\bf i}\sigma}$) denotes the density of charge carriers 
having spin $\sigma$ at molecular sites ${\bf i}$. The electrons are assumed to be coupled 
to the intra-molecular deformations $u_{\bf i}$ via  charge density fluctuations and the 
coupling constant is denoted by $\lambda$. The dynamics of the 
lattice is treated purely locally with Einstein oscillators describing 
the intra-molecular oscillations with frequency $\omega_0$. $M$ denotes 
the mass of the atoms making up the diatomic molecular units. The 
additional Hubbard $U$ intra-molecular repulsion is introduced sometimes
in order to account for possible correlation effects in conjunction with 
the purely polaronic features. We shall not consider the effect of this term 
in this present discussion, but several specific lectures in this school  will be 
devoted to it.

This model is capable of describing the self-trapping of charge carriers 
which arises from a competition between the energy gain coming from the 
itinerancy of the electrons and  that coming from the potential energy 
due to the induced local deformations of the molecular units. In order to 
see that let us rewrite this Hamiltonian in a form which makes more 
explicit such a  self-trapped localized picture:
\begin{eqnarray}
H_0 &=& (D-\varepsilon_p)\sum_{\bf i}n_{\bf i} -  t\sum_{{\bf i}\neq 
j\sigma}(c^+_{{\bf i}\sigma}c^{\phantom{+}}_{{\bf j}\sigma}+
h.c.)+
\sum_{\bf i}(U-2\varepsilon_p)n_{{\bf i}\uparrow}n_{{\bf i}\downarrow} \nonumber \\  
&&+ \sum_{\bf i}\frac{M}{2}\left[\left(\dot{u_{\bf i}}-\frac{\lambda \dot{n_{\bf i}}}
{M\omega_0^2}\right)^2+
\omega^2_0\left(u_{\bf i}-\frac{\lambda n_{\bf i}}{M\omega_0^2}\right)^2\right]
- \frac{M}{2}\left[\left(\frac{\lambda \dot n_{\bf i}}{M \omega_0^2}\right)^2-
2\dot u_{\bf i}\frac{\lambda \dot n_{\bf i}}{M \omega_0^2}\right].
\label{eq28}
\end{eqnarray}
The major features which evolve out of such a representation are:
\begin{itemize}
\item   
The energy of the electrons is lowered by an amount 
$\varepsilon_p=\lambda^2/2M\omega^2_0$ which corresponds to the 
ionization energy of the polaron
\item
A lattice induced intra-molecular attraction of strength $2\varepsilon_p$ between 
the electrons on different sites of the diatomic molecule units  which can partially or 
totally compensate their intra-molecular Coulomb repulsion $U$.
\item
The intra-molecular distance of the Einstein oscillators are shifted by a time dependent 
quantity $u^0_{\bf i}$~=$\lambda n_{\bf i}/M\omega_0^2$ which follows the time evolution 
of the charge redistribution on site ${\bf i}$. The frequency of the oscillators 
is  modified as a consequence.
\end{itemize}

Since in this school we shall primarily be concerned with systems which can 
be described by the Holstein model, let us now focus in more detail on the 
basic physics inherent in this scenario. To begin with, we consider this 
problem in a semi-classical fashion and in the adiabatic approach. 

In the limit of small coupling $\varepsilon_p\ll D$, the form of the Hamiltonian, given in eq.~(\ref{eq28}) is reminiscent of one which describes itinerant electrons in a static potential, given by $V^{\rm wc}(\{u_i\})$~= $-\lambda \langle u_i\rangle c^+_{{\bf i}\sigma} c^{\phantom +}_{{\bf i}\sigma} d^+_{\bf i}d^{\phantom +}_{\bf i}$ with  $\langle u_{\bf i}\rangle$~= $\lambda/M\omega^2_0$ and $d^+_{\bf i}d^{\phantom +}_{\bf i}$ representing the density of 
some fictitious particles. Solving the eigenvalue problem for the ground state 
energy $E_0$, i.e., 
\begin {equation}
{\lambda^2 \over M \omega^2_0}\sum_{\bf k} {1 \over \varepsilon_{\bf k} - E_0} = 1
\label{eq29}
\end{equation}
predicts a splitting off of the ground state energy from the bottom of the free 
itinerant band ($E_0<0$), resulting in a localization of the charge carriers.

\begin{figure}
\begin{minipage}[c]{6.5cm}
\includegraphics[width=6.5cm]{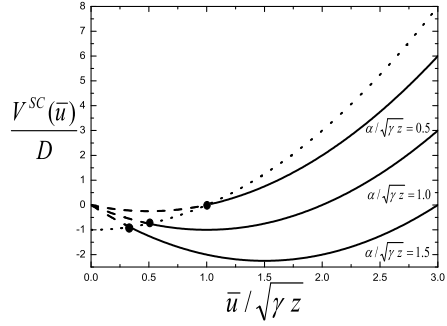}
\end{minipage}
\hspace*{0.15cm}
\begin{minipage}[c]{6.75cm}
\caption{The variation of $V^{\rm sc}(\tilde{u})$ 
(full lines) felt by an electron with wave vector $k$=0 in the presence of a static continuous deformation of the intramolecular distance 
$\bar{u}$ for various values of $\alpha$ and $\gamma$. The dotted line describes the  energy of the electron with homogeneously deformed oscillators uncoupled to the electron. The crossing points (dots) of these curves indicate a change from itinerant to localized behavior.}
\label{V}
\end{minipage}
\end{figure}

In the limit of strong coupling, for large values for $\lambda$ such that  
$\varepsilon_p \geq  D$,  the adiabatic approach tells us to ignore the 
electron itinerancy  and assume the electron to be fixed at a particular lattice 
site. This then results in an adiabatic potential for the electron given by 
$V^{\rm sc}(\{u_{\bf i}\})$~= $-\varepsilon_{\rm p}n_{\bf i}+\frac{M}{2}\omega^2_0\left(u_{\bf i}-\frac{\lambda n_{\bf i}}{M\omega_0^2}\right)^2$, which follows directly from the form of the Hamiltonian given in eq.~(\ref{eq28}). If one requires that this potential is deep enough to bind the electron in the first place, the selfconsistency of the adiabatic approach is guaranteed and leads to localized states in this strong coupling limit where the time derivatives of the local deformations can be neglected.  As $\lambda$ is varied, these potentials for the weak and strong coupling limits are expected to join up smoothly. Depending on the strength of the coupling  $\lambda$,  this small exercise shows that 
one can have a situation where the energies of the two configurations are degenerate. In this intermediary coupling regime  one should expect a coexistence of small localized polarons and weakly bound electrons. It has become customary in the theory of small polarons to introduce two dimensionless parameters: $\alpha\equiv \sqrt{\varepsilon_p/\hbar \omega_0}$ measuring the strength of the interaction and $\gamma$~= $t / \hbar \omega_0$  the adiabaticity ratio ($\alpha$ has nothing to do with the dimensionless coupling constant 
$\alpha_{\rm Fr}$ used in the theory of large polarons and introduced in section \ref{FLP}). We then can rewrite the potential $V^{\rm sc}(\{u_i\})$ in a compact form in terms of those parameters and a dimensionless local lattice deformation $\bar{u}_{\bf i}$~= $u_{\bf i} \sqrt{M\omega_0/2\hbar}$. Its variation with $\alpha$ and $\gamma$ as a function of $\bar{u}$  illustrated in fig.~\ref{V}.

This picture of the polaron physics, widely used and appreciated until the nineteen sixties (see ref.~\cite{polarons_excitons63}), although intuitively very appealing is qualitatively incorrect - as we shall see in the next 
section. There we shall show that a polaron, although self-trapped, remains delocalized for any dimension.

A formal and correct treatment of this problem was first initiated  by 
Holstein and collaborators \cite{HolsteinAP59,EminAP69} and in a very elegant and 
efficient way in terms of a unitary transformation by Firsov and 
collaborators \cite{LangJETP63}, given by
\begin{equation}
\tilde H = e^S H e^{-S}, \qquad S = \alpha \sum_{{\bf i}\sigma} 
n_{{\bf i}\sigma}(a_{\bf i} - a^+_{\bf i})\, ,
\label{eq30}
\end{equation}
which transform the electron operators $c^{(+)}_{{\bf i}\sigma}$ into operators which describe charge carriers rigidly  tied to local lattice deformations
\begin{equation}
\tilde c^+_{{\bf i}\sigma}  =  c^+_{{\bf i}\sigma} X^+_{\bf i}, \qquad   \tilde c_{{\bf i}\sigma}  =  
c_{{\bf i}\sigma} X^-_{\bf i}, \qquad
X^{\pm}_{\bf i} = e^{\pm \alpha(a_{\bf i} - a^+_{\bf i})}
\label{eq31}
\end{equation}
and which correspond to localized polaronic self-trapped states
\begin{equation}
X^+_{\bf i}|0)_{\bf i} = \sum_n e^{-\frac{1}{2}\alpha^2}{(\alpha)^n \over \sqrt{n!}} |n)_{\bf i}\, .
\label{eq32}
\end{equation}
$|n)_{\bf i}$ denotes the n-th excited oscillator states at a molecular sites ${\bf i}$ and 
$X^+_{\bf i}|0)_i$ signifies an   oscillator ground state  whose equilibrium position is  
shifted by an amount  $\lambda/M \omega^2_0$. The corresponding transformed Hamiltonian is consequently given by 
\begin{eqnarray}
\tilde H &=& \sum_{{\bf i}\sigma} (D - \varepsilon_p) c^+_{{\bf i}\sigma} c_{{\bf i}\sigma}  - 
t\sum_{{\bf i} \neq {\bf j}\sigma} (c^+_{{\bf i}\sigma} c_{{\bf j}\sigma} X^+_{\bf i}X^-_{\bf j} +  H.c.)\nonumber\\
&&+ \sum_{\bf i}(U - 2 \varepsilon_p)c^+_{{\bf i}\uparrow} c^+_{{\bf i}\downarrow} c_{{\bf i}\downarrow}c_{{\bf i}\uparrow} + \hbar \omega_0 \sum_{\bf i}(a^+_{\bf i} a_{\bf i} + \frac{1}{2})\,.
\label{eq33}
\end{eqnarray}

Such an approach to the polaron problem is particularly useful in the limit of 
large $\alpha$ where $\varepsilon_p   \geq D$ and for a small adiabaticity ratio $\gamma$. In the strong coupling limit and anti-adiabaticity ($\gamma < 1$), the term  $X^+_{\bf i}X^-_{\bf j}$ in the transformed Hamiltonian can be averaged over the bare phonon states which,  
for zero temperature, results in an effective polaron hopping integral 
$t^* = t e^{-\alpha^2}$. This is justified a posteriori since then $t^*\ll\omega_0$, where the local lattice deformations can be considered to adapt themselves quasi instantaneously to the slowly in time varying positions of the electrons. Such an  approximation implies that the number of phonons in the phonon clouds surrounding each charge carrier remains largely unchanged during the transfer of a charge carrier from one site to the next, while processes where the number of phonons in the clouds change give rise to a polaron damping. Neglecting 
such damping effects, one obtains well defined Bloch states for the polaronic charge carriers (defined by $c^+_{{\bf i}\sigma} X^+_{\bf i}$) , albeit with a much reduced hopping integral $t^*$, while at the same time 
the electrons loose  practically  all of their coherence and thus their 
quasi-particle features. In order to illustrate that, let us consider the Green's 
function for a single localized  polaron,  respectively for a 
single localized  electron by putting $t^* = 0$. For the polaron retarded Green's function we have
\begin{eqnarray}
G_{\rm p}^{\rm ret}(t) &=& -\theta(t)(0|X^-_{\bf i}[ e^{{\bf i} \tilde H t} \tilde{c}^+_{{\bf i}\sigma}(0) 
e^{-{\bf i} \tilde H t}\tilde{c}_{{\bf i}\sigma}(0)] X^+_{\bf i}|0)\nonumber\\
 &=& -\theta(t)\sum_n e^{{\bf i}( \varepsilon_{\rm p} - n\hbar \omega_0)t}
|(0|X^-_{\bf i}  \tilde{c}^+_{{\bf i}\sigma}(0) |n)|^2 \delta_{n,0}=-\theta(t) 
e^{{\bf i}\varepsilon_{\rm p}t}|(0|c^+_{{\bf i}\sigma}(0) |0)|^2
\label{eq34}
\end{eqnarray}
which after Fourier transforming becomes 
\begin{equation}
G_{\rm p}^{\rm ret}(\omega) = \lim_{\delta \rightarrow 0}
{1  \over \omega + i\delta + \varepsilon_{\rm p}}
\label{eq35}
\end{equation}
and displays a spectrum which consists of exclusively a coherent 
contribution. On the contrary, the electron retarded Green's function
\begin{eqnarray}
G_{\rm el}^{\rm ret}(t) &=& -\theta(t)(0|X^-_{\bf i}[ e^{i \tilde H t} c^+_{\bf i}(0) 
e^{- i \tilde H t}c_{{\bf i}\sigma}(0)] X^+_{\bf i}|0)\nonumber\\
 &=& \sum^{\infty}_{n=0} e^{i( \varepsilon_{\rm p} - n\hbar \omega_0)t}
|(0|X^-_{\bf i}  c^+_{{\bf i}\sigma}(0) |n)|^2
\label{eq36}
\end{eqnarray}
displays a spectrum given by its Fourier transform
\begin{equation}
G_{\rm el}^{\rm ret}(\omega)\simeq \lim_{\delta \rightarrow 0}\sum^{\infty}_{n=0}
{e^{-\alpha^2}\alpha^{2n} 
\over n!} {1  \over \omega + i\delta + \varepsilon_{\rm p} - n \hbar \omega_0}\, .
\label{eq37}
\end{equation}
The spectral weight of the coherent part is now reduced 
to $\exp(-\alpha^2)$, corresponding to the term $n$~= 0 in eq.~(\ref{eq37}), while the major part of the spectrum is made up by the incoherent contributions which track the structure of the composite nature of the polaron.

Generalizing these results by including the itinerancy of the charge carriers 
leads to very similar results \cite{RanningerPRB93} for the electron Green's 
function for a many polaron system, which in this strong coupling anti-adiabatic limit reduces to a system of small polarons in band states and where the Green's function for the electrons  with wavevectors ${\bf k}$ is given by

\vbox{\begin{eqnarray}
G_{\rm el}^{\rm ret}({\bf k},\omega) &=& \lim_{\delta \rightarrow 0}
{e^{-\alpha^2} \over \omega +i\delta +\varepsilon_{\rm p} - 
{\varepsilon}^*_{\bf k}} 
\nonumber \\
&+& \lim_{\delta \rightarrow 0}e^{-\alpha^2}
\sum^{\infty}_{n=1}{\alpha^{2n} \over n!}{1 \over N}\sum_{{\bf k}'}
\left[{f({\varepsilon}^*_{{\bf k}'}) 
\over \omega + i\delta - {\varepsilon}^*_{{\bf k}'} + n \hbar \omega_0} + 
{1 - f({\varepsilon}^*_{{\bf k}'}) \over \omega + i\delta - 
{\varepsilon}^*_{{\bf k}'} - 
n \hbar \omega_0}\right] 
\label{eq38}
\end{eqnarray}}
\noindent and where  $\varepsilon^*_{\bf k}$~= $e^{-\alpha^2}\varepsilon_{\bf k}$.
These characteristic spectral properties of small polarons can be tested by 
photo-emission spectroscopy and present crucial tests which permit to distinguishing between different mechanisms leading to heavily dressed composite quasi-particles. The Poisson type phonon distribution was particularly well demonstrated early on in such experiments on simple molecules such as $H_2$ molecules \cite{AsbrinkCPL70} and carbon rings \cite{HandschuhPRL95}, on localized polaron states in the 
manganites by neutron spectroscopy \cite{LoucaPRB99} and in the cuprate superconductors by infrared absorption measurements \cite{BiPRL93}.

\begin{figure}
\begin{minipage}[c]{6.5cm}
\includegraphics[width=6.5cm]{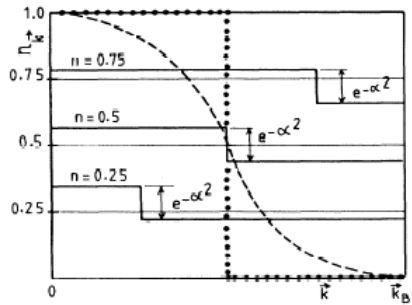}
\end{minipage}
\hspace*{0.15cm}
\begin{minipage}[c]{6.75cm}
\caption{Occupation number $n_{{\bf k}\sigma}$
for a system of itinerant small polarons for  $\alpha\gg 1$. The dashed and 
dotted curves indicate the case $n = 1$ for  $\alpha \simeq 1$ and  
$\alpha\ll 1$ respectively (after ref.~\cite{RanningerPRB93}).}
\label{n-k}
\end{minipage}
\end{figure}

The localized, or almost localized, nature of such small polarons is apparent in their 
electron occupation number distribution illustrated in fig.~\ref{n-k} 

showing an almost flat distribution covering the whole Brillouin zone, i.e.,
\begin{equation}
n_{{\bf k}\sigma} = \langle c^+_{{\bf k}\sigma}c^{\phantom +}_{{\bf k}\sigma}\rangle = 
(1 -e^{-\alpha^2})n_{\sigma} + e^{-\alpha^2}n_{\sigma}(\varepsilon^*_{\bf k})\, .
\label{eq39}
\end{equation}

There is a qualitative difference between the weak coupling and the strong 
coupling limit discussed here. We have seen that in the weak coupling limit of the Fr\"ohlich Hamiltonian the electrons are accompanied by  phonons with which they continuously exchange the momentum such as to keep the total momentum of electron plus phonon constant. In the strong coupling limit the electrons are surrounded by real lattice deformations which in terms of phonons means clouds of phonons and which in the limit of extreme strong coupling, where the electrons can be considered as localized on a given site, have a Poisson distribution. The new feature now is that this deformation corresponds to a local mass inhomogeneity capable of  carrying true momentum rather than the pseudo-momentum as is the case for weak coupling. Again, as we shall see below, the momentum between the electron and this deformation is perpetually exchanged as the polaronic charge carrier moves through the lattice, leading to a space-time dependent interaction between the charge carrier and the deformation associated to it. This effect will be particularly important in the cross-over between the weak coupling adiabatic and the strong coupling anti-adiabatic regime, discussed in the next section.

A key point which occupied the field of polaron theory for several decades was to try to establish if a large delocalized polaron changes discontinuously or continuously---although abruptly--- into a small self-trapped polaron as the electron-lattice coupling is increased beyond a certain critical value. We shall review this issue in the following.

\section{Self-trapping}

We shall in this section discuss the meaning of self-trapping. For that purpose let us go back for a moment to the semi-classical approach of the polaron problem  (discussed in section \ref{Hsp} above) and the adiabatic lattice potential which goes with it. On the basis of such an approach, amounting to treat the polaron problem as a potential problem, it is tempting to conclude localization of charge carriers as their coupling to the lattice degrees of freedom increases. As we shall see below, the polaron problem can not be treated as a potential problem. Its intrinsic dynamics of the local lattice deformations is determinant in correlating the dynamics of the charge and the lattice degrees of freedom, thus resulting in itinerant delocalized states of electrons surrounded by clouds of phonons whose density varies and increases as we go from the adiabatic weak 
coupling to the anti-adiabatic strong coupling limit. For that reason the lattice 
degrees of freedom must be treated in a quantized version.

Let us start with the weak coupling 
limit for which we can assume a  polaron state in the form 
\begin{equation}
|\Psi\rangle_{\bf k} = {1 \over  \sqrt N} \sum_{\bf ij}e^{{\bf ik} \cdot {\bf r}_{\bf i}}
[c^+_{{\bf i}\sigma}\delta_{\bf ij} +  f_{\bf ij}^{\bf k}c^+_{{\bf j}\sigma} a^+_{\bf i}]|0\rangle\, ,
\label{eq40}
\end{equation}
which is the real space version of the state previously discussed, i.e., eq.~(\ref{eq24}). Diagonalizing the Hamiltonian within such a subspace of zero and respectively one phonon present, leads to the following selfconsistent equations for the eigenvalues and parameters $f_{ij}^{\bf k}$.
\begin{eqnarray}
E_{\bf k} &=& \varepsilon_{\bf k} - (\alpha \hbar \omega_0)^2 \frac{1}{N}\sum_{{\bf k}'} 
{1 \over \varepsilon_{{\bf k} - {\bf k}'} - E_{\bf k} + \hbar \omega_0}\,,\label{eq41} \\
f^{\bf k}_{ij} &=& \alpha \hbar \omega_0 \frac{1}{N}\sum_{{\bf k}'} 
{e^{i {\bf k}' \cdot ({\bf r}_i - {\bf r}_j)} \over 
\varepsilon_{{\bf k} - {\bf k}'} -E_{\bf k} + \hbar \omega_0}\, .\label{eq42}
\end{eqnarray}

Considering the solution for the ground state (${\bf k}=0$) of this problem, one  
notices that its corresponding eigenvalue $E_0$ falls below the bottom of the free electron band ($E_0 \leq 0$). This signals a bound state which is characterized by an exponential drop-off of the spatial correlation between the electron and the accompanying phonon, which for a 3D system is given by:  
\begin{equation}
f^0_{\bf ij} = v\frac{\alpha \omega_0 m}{2 \pi \hbar}\frac{1}
{|{\bf r}_{\bf i}- {\bf r}_{\bf j}|}e^{-\sqrt{2m(-E_0 +\hbar \omega_0)}
\frac{|{\bf r}_{\bf i}- {\bf r}_{\bf j}|}{\hbar}}
\label{eq43}
\end{equation} 
($v$ denoting the unit cell volume) and which indicates that the dynamically 
(not thermally!) excited phonons accompanying the electron remain in its immediate vicinity. The exponential drop-off does not depend  on any dimensionality (in contrast to the semi-classical approach outlined in section \ref{Hsp}) and gives us a first indication for the intricate correlation between the inherent dynamics of lattice fluctuations and the charge dynamics. Hence, already in this weak coupling limit, the polaron problem can not be reduced to an effective (adiabatic) potential problem.

\begin{figure}
 \begin{minipage}[b]{5.5cm}
\includegraphics[width=5.5cm]{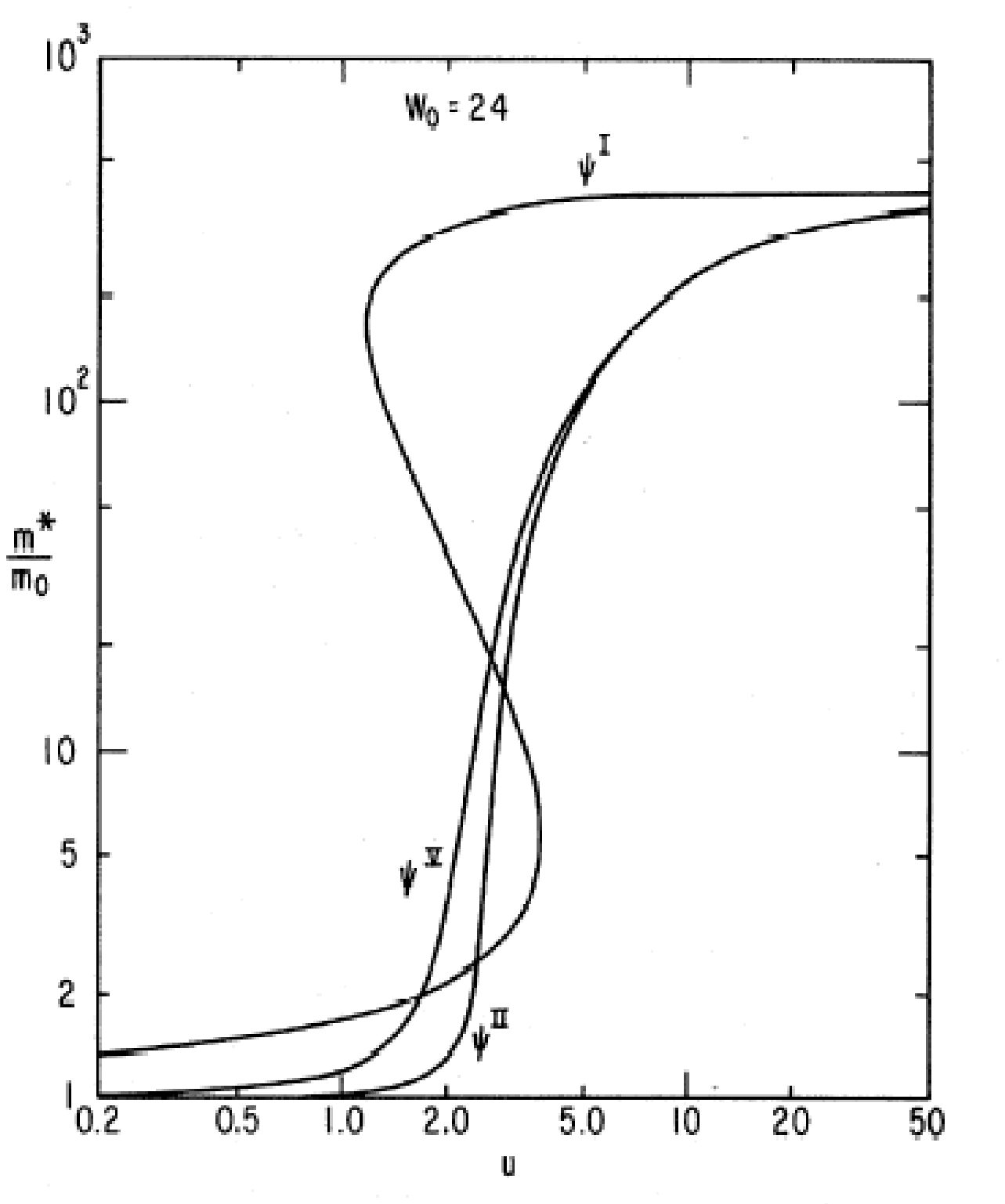}
\caption{The effective mass versus the electron-lattice coupling $u$ 
(= $\alpha^2(\gamma 2z)^{-1}$ in our notation) , after ref.~\cite{ShorePRB73}, for oscillator trial wave functions $\Psi^{I}$ and  $\Psi^{II}$ such as given by  eqs.~(\ref{eq45}), (\ref{eq46}).}
\label{Shore1}
 \end{minipage}
 \hspace*{0.25cm}
 \begin{minipage}[b]{7.5cm}
\includegraphics[width=7.5cm]{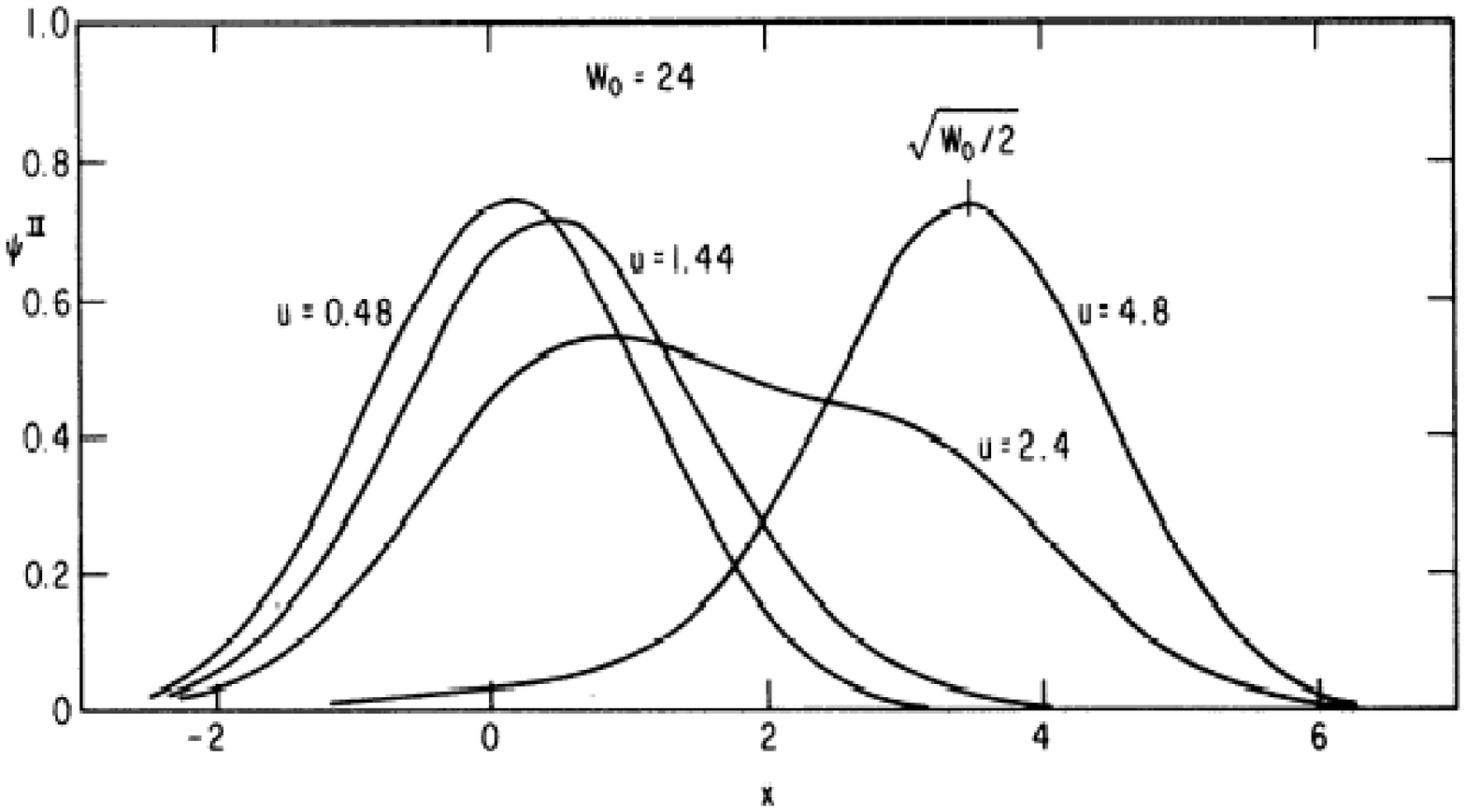}
\caption{The variation of the oscillator wave function on a given site (after ref.~\cite{ShorePRB73}) as a function of $ \{u,\,W_0\}$ (= $\{ \alpha^2(\gamma 2z)^{-1}),\,\alpha^2 \}$ in our notation) for oscillator trial wave functions $\Psi^{II}$ such as given by eq.~(\ref{eq46}) and with $x = \frac{1}{\sqrt 2}\langle a^+_i + a_i\rangle$.}
\label{Shore2}
 \end{minipage}
\end{figure}

Let us next turn to the strong coupling limit, where on the basis of such 
an adiabatic approach one would expect self-trapped lowest energy eigenstates of the form
\begin{equation}
|\Psi\rangle_{\bf k} = {1 \over \sqrt N}\sum_{\bf i} e^{i{\bf k}\cdot{\bf r}_i}c^+_{\bf k}
|\Phi_{\bf k}(\{u_j\})\rangle\, ,
\label{eq44}
\end{equation} 
where $\Phi_{\bf k}(\{u_{\bf j}\})$ denotes a function  of the ensemble of oscillator 
coordinates ${u_{\bf j}}$ situated at sites ${\bf j}$ in the vicinity of site ${\bf i}$ where the 
electron is located. Guided by the semi-classical picture of the adiabatic 
potential, resulting in the  strong coupling limit, one can try an intuitively 
appealing Ansatz for the lattice wavefunction in form of a series of 
displaced oscillators around the site where the electron is situated and 
which, for the ground state, can be assumed to be of the form
\begin{equation}
|\Phi_{{\bf k}=0}(\{u_{\bf j}\})\rangle= e^{-\alpha
\sum_{\bf j}(f^*_{\bf ij} a^+_{\bf j}- f_{\bf ij} a_{\bf j})}|0\rangle \, .
\label{eq45}
\end{equation} 
Determining the parameters $f_{\bf ij}$ variationally \cite{ToyozawaPTP61} leads to solutions which change discontinuously in the intermediary coupling regime. 
Associated with that is a discontinuous change of the mass of the corresponding quasi-particles amounting to a change-over from a practically free band dispersion with a electron mass $m_{\rm el}$ to an effective polaron mass $m_{\rm p}$ given by  $m_{\rm p}/m_{\rm el}=e^{\alpha^2}$ (see fig.~\ref{Shore1}). 

It has been a matter of dispute for many decades and up to the 1960-ties whether this discontinuity is a real effect or is an artifact of the 
theory. From the experimental side, small polarons where found to be localized 
and have ever since, for the presently available materials,  shown a mobility 
via hopping  rather than Bloch like band motion and metallic conductivity.
The theoretical results obtained for the cross-over regime between large and small polarons on the basis of the semi-classical description in terms of an adiabatic lattice potential with two minima of comparable energy (see the discussion in section \ref{Hsp}), suggest already that in this regime one should have particularly strong  fluctuations of the local lattice displacements which could possibly be modeled by some effective lattice wave function  in terms of a superposition of two sorts of oscillator states: one practically undisplaced and one displaced oscillator state on  the  site housing the electron, as hypothesized early on in a slightly different context \cite{EaglesPSSB71}. These ideas were subsequently followed up by a variety of different approaches such as: exact diagonalization studies \cite{ShorePRB73} 
on small clusters, upon assuming the charge induced deformation to be constraint to the site where the electron sits \cite{ChoJPSJ71} (when the problem can be solved analytically) and Quantum Monte Carlo studies \cite{deRaedtPRB83}.

All these approaches converge to a lattice wave function for the ground state  
which can be approximated by
\begin{equation}
|\Phi_{\bf p}(\{u_{\bf j}\})\rangle=
[e^{-\alpha\sum_{\bf j}(f^*_{\bf ij}a^+_{\bf j}- f_{\bf ij}a_{\bf j})} + 
\eta e^{-\alpha
\sum_{\bf j}(g^*_{\bf ij}a^+_{\bf j}- g_{\bf ij} a_{\bf j})}|]0\rangle \, .
\label{eq46}
\end{equation} 
This results (after variationally determining the parameters $f$, $g$ and $\eta$) 
in a smooth, but nevertheless very abrupt, cross-over between 
the weak and strong coupling features and a change-over from a quasi-free electron band 
mass to a strongly enhanced one (see fig.~\ref{Shore1}). The form of eq.~(\ref{eq46}) represents for any electron situated on a given site  to be associated with an oscillator 
wave function on this site, illustrated in fig.~\ref{Shore2}. It indicates an essentially 
undisplaced oscillator for weak coupling, a displaced one for strong coupling and a 
superposition of such two oscillator states for the cross-over regime between those two 
limits. This latter has led to the suggestion of  strong 
retardation effects between the dynamics of the electron and that of the local lattice 
deformation which could possibly result in a dynamically disordered system and 
subsequent localization, a scenario which we shall return to in the lecture ``From Cooper-pairs to resonating Bipolarons''. 

Let us here consider this point in more detail and investigate this cross-over behavior 
in terms  of the  strong dynamical correlation which act between the charge and the lattice 
degrees of freedom. We shall
demonstrate that on hand of a polaron toy problem, such as a two-site system involving 
two adjacent diatomic molecules whose individual oscillations are uncorrelated with 
each other and an electron with spin $\sigma$  hopping between those two molecular 
units. The Hamiltonian, eq.~(\ref{eq27}), for that then reduces to
\begin{eqnarray}
H &=& t(n_{1\sigma} +n_{2\sigma})-
t(c^{+}_{1\sigma}c^{\phantom{+}}_{2\sigma} + 
c^{+}_{2\sigma}c^{\phantom{+}}_{1\sigma})
-\lambda(n_{1}u_1+n_{2}u_2) \nonumber\\
&&+ \frac{M}{2} \left[(\dot{u}^2_1+ \dot{u}^2_2)  + \omega^2_0(u^2_1+u^2_2)\right]\, .
\label{eq47}
\end{eqnarray}
Upon introducing the variables 
\begin{equation}
X ={u_1 - u_2 \over \sqrt{2}},\qquad Y ={u_1 + u_2 \over \sqrt{2}}
\label{eq48}
\end{equation}
this  Hamiltonian separates into two contributions $H_X$ and $H_Y$ depending 
respectively on lattice coordinates  $X$ and $Y$. $Y$ couples exclusively 
to the total number of charge carriers 
in the system, ie. $(n_{1\sigma} + n_{2\sigma})$. It presents an 
in-phase oscillation of the two molecules and has hence no effect on the dynamics 
of the polaron and will be disregarded for the present considerations. The lattice 
coordinate $X$, on the contrary, presents an out of phase oscillation of the 
two molecules and couples to the relative charge distribution  
$(n_{1\sigma} - n_{2\sigma})$. The Hamiltonian for this subspace 
which describes the intertwined dynamics of the charge and the lattice 
deformations is 
 \begin{equation}
\quad H_X = t(n_{1\sigma} + n_{2\sigma})-
t(c^{+}_{1\sigma}c^{\phantom{+}}_{2\sigma} + 
c^{+}_{2\sigma}c^{\phantom{+}}_{1\sigma})
-{\lambda \over \sqrt{2}}  (n_{1\sigma} -  n_{2\sigma}) X  
 + \frac{M}{2} \left[\dot{X}^2  +  \omega^2_0 X^2\right]\, . 
 \label{eq49}
\end{equation}
The term proportional to $\lambda$ in this Hamiltonian clearly indicates 
how the charge fluctuations between the two molecules induces a 
coupling between the oscillators of those two molecules and which ultimately couples 
back to the dynamics of the charge transfer. This generates, as we shall see, a 
\underline{new} 
dynamical lattice deformation mode which accompanies the transfer of charge from one 
to the other molecule in a slow and smooth continuous fashion. Upon introducing the 
quantification of the out of phase lattice coordinate $X=(a^{\phantom{+}}+a^+)/\sqrt{2M\omega_0/\hbar}$ in terms of corresponding phonon operators $a^+, a^{\phantom{+}}$ and diagonalizing the above Hamiltonian in a truncated Hilbert space (keeping the 
number of phonon states finite but large) \cite{RanningerPRB92} we obtain for the 
ground state of this system with a single electron with spin $\sigma$:
\begin{equation}
|GS\rangle_{\sigma} = \frac{1}{\sqrt 2}\left[c^+_{1\sigma} |\psi^0_+(X)\rangle +  
c^+_{2\sigma} |\psi^0_-(X)\rangle\right]\, .
 \label{eq51}
\end{equation}
$|\psi^0_{\pm}(X)\rangle$ denote oscillator states in real space 
when the electron is either situated on site 1 or site 2, similar to the oscillator 
wave functions illustrated in fig.~\ref{Shore2}. In particular, in the cross-over regime,
these oscillator wave-functions reflect the bimodal probability distribution of the 
equilibrium  positions of the oscillator where the electron sits. Let us next investigate 
the  evolution in time of the charge transfer together with that of the inter-molecular 
deformation, given by the correlation functions:
 \vbox{
\begin{equation}
\chi_{_{nn}}(\tau) = \theta(\tau)\langle [n_{1\sigma}(\tau) - n_{2\sigma}(\tau)]
[n_{1\sigma}(0) - n_{2\sigma}(0)] \rangle
 \label{eq52}
 \end{equation}
 \begin{equation}
\chi_{_{XX}} = \theta(\tau)\langle X(\tau) X(0)\rangle\, .
 \label{eq53}
\end{equation}}

\begin{figure}
 \begin{minipage}[t]{6.5cm}
\includegraphics[width=6.5cm]{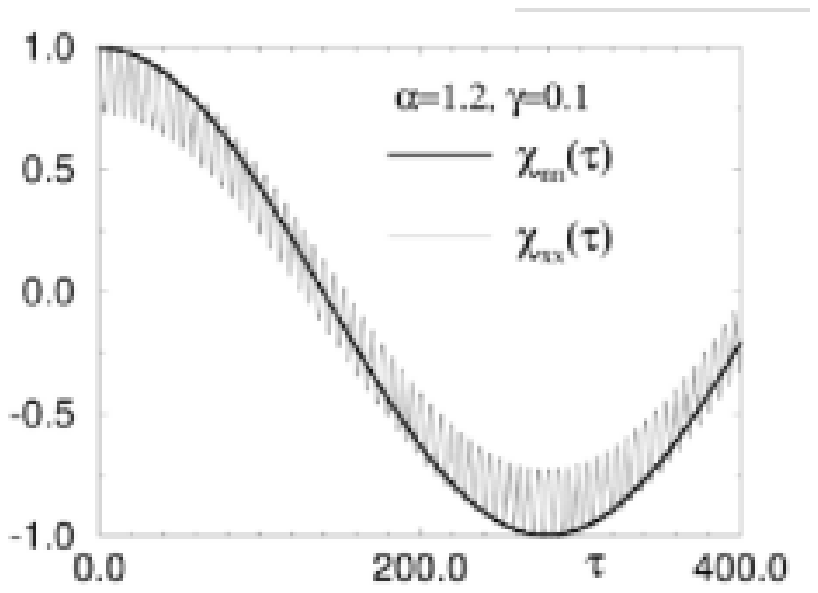}
\label{DynCor1}
 \end{minipage}
 \hspace*{0.25cm}
 \begin{minipage}[t]{6.5cm}
\includegraphics[width=6.5cm]{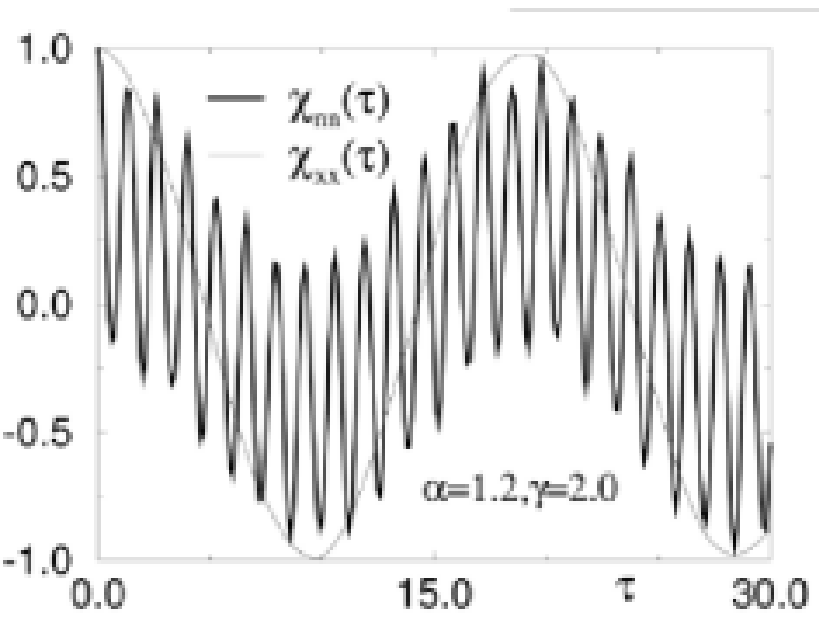}
\label{DynCor2}
 \end{minipage}

\begin{minipage}[c]{6.5cm}
\includegraphics[width=6.5cm]{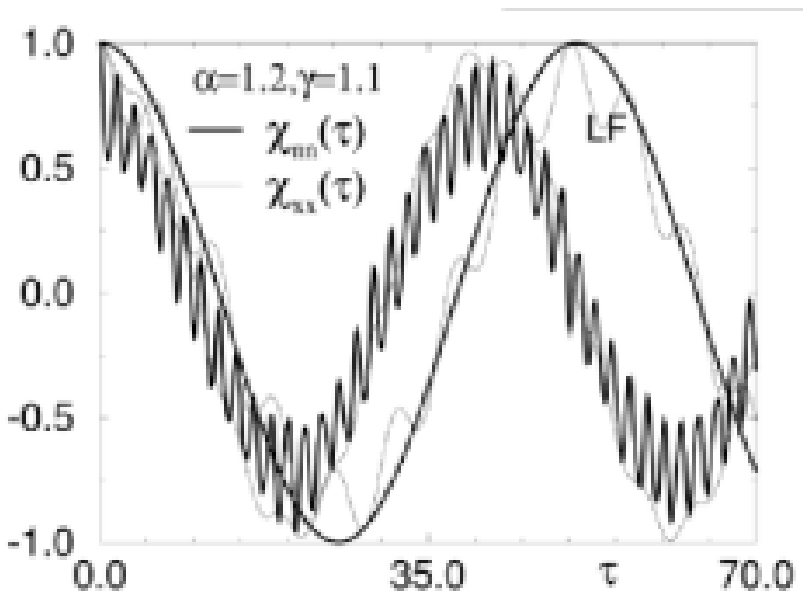}
 \end{minipage}
 \hspace*{0.15cm}
 \begin{minipage}[c]{6.75cm}
\caption{The evolution in time (in units of $\omega^{-1}_0$) of the charge and intra-molecular distance transfer between two adjacent molecular units for a two-site one-electron system (after ref.~\cite{deMelloPRB97}) in the anti-adiabatic ($\gamma$~= 0.1), the adiabatic ($\gamma$~= 2.0) and in the cross-over regime ($\gamma$~= 1.1) and for (in the present notation) a coupling strength $\alpha=1.2\sqrt{2}\simeq 1.7$.}
\label{DynCor3}
 \end{minipage}
\end{figure}

The results of this are reproduced in fig.~\ref{DynCor3} 
for a fixed value of $\alpha$ (which in the present notation corresponds to 
$\alpha = \sqrt 2 \cdot 1.2$~= 1.70) and where we cover the cross-over from 
the strong coupling anti-adiabatic limit ($\gamma$~= 0.1) to the strong coupling 
adiabatic limit ($\gamma= 2.0$) upon increasing the adiabaticity ratio. 

The anti-adiabatic limit is characterized by a smooth and slowly 
in time varying transfer of charge from one molecule to the next. This charge transfer 
then {\it slaves} the inter-molecular deformation $X$ by subjecting it to a 
slowly in time varying driving force 
$-{\lambda \over \sqrt{2}} [n_{1\sigma}(\tau) -  n_{2\sigma}(\tau)]X(\tau)$. This in turn leads to a slowly in time varying sinusoidal intermolecular deformation, onto 
which are superposed the local intrinsic oscillations of the individual molecules 
with a frequency of the order of $\omega_0$. This is the exact opposite of what 
happens in the extreme other limit, when we approach the adiabatic situation.

In this adiabatic regime,  considering here  an 
adiabaticity ratio $\gamma=2.0$, the slowly in time varying quantity is now 
the inter-molecular deformation (showing almost no fluctuations of the 
individual molecules with frequency $\omega_0$), which {\it slaves} the intermolecular 
charge transfer characterized by  $[n_{1\sigma}(\tau) -  n_{2\sigma}(\tau)]$ . Averaged 
over several periods of oscillations of frequency $\simeq 1/t$,  this charge transfer  
adiabatically follows again the slowly in time varying  potential 
$-{\lambda \over \sqrt{2}} [n_{1\sigma}(\tau)-  n_{2\sigma}(\tau)] X(\tau)$. 

These results show that in these two extreme limits of adiabaticity and anti-adiabaticity 
the slowly in time varying component of the charge transfer 
processes and of the inter-molecular deformation dynamics are completely locked 
together. This is no longer the case in the cross-over regime between these two extreme 
limits, here shown in fig.~\ref{DynCor3} for the adiabaticity ratio $1.1$. 
One can clearly distinguish phase-slips occurring between the correlated motion of 
the charge and the deformation transfer, which, in a large system where many body 
effects become important, could possibly lead to a dynamically induced localization 
of the charge carriers. 

\section{Some early experimental insights}\label{earlyexperinsights}

The local character of small polaronic charge carriers requires specific experimental probes able to track the polaron induced lattice as well as charge excitations on a short spatial ($\simeq$~5~\AA) as well as time ($10^{-13}-10^{-15}$~sec) scale. Generally such probes are quite adequate for insulating polaronic systems with either charge ordered bipolarons (Ti$_4$O$_7$), bipolaronic Mott type insulators (WO$_{3-x}$, the manganites and nickelates) and low density polaron systems arising from photo-induced or chemical doping and situated on the verge of a quantum phase insulator-superconductor transition (high $T_c$ cuprates). Concerning the cuprates, manganites and nickelates we refer the reader to the lectures by T.~Egami and N.~L.~Saini in this volume, dealing with pulsed inelastic neutron scattering techniques and EXAFS as well as XANES. Here we shall restrict ourselves to the discussion of more classical probes such as optical absorption which can select specific local polaron-sensitive lattice modes.

As we have seen in the discussion presented above, one of the major issue here is to track the disintegration and reconstruction of a polaron during its transfer from one site to the next. This process implies a gradual stripping off of the electron's phonon cloud upon leaving a given site and a subsequent rebuilding of this phonon cloud on the new site where the electron eventually ends up. Another important issue in the polaron problem is connected to the physics of Many-Polaron system, caused by the polaron induced local attraction which can bind two polarons on a given effective site into a bipolaron. We shall now discuss specific experiments which can illustrate these two characteristic polaron features.

Shortly after Anderson's suggestion \cite{AndersonPRL75} of localized bipolarons 
in amorphous chalcogenide glasses, exhibiting a natural diamagnetism resulting from  
covalent bonding in locally deformed lattice structures, a variety of systems 
were found where bipolarons existed in more or less dense situations. Examples for dense 
bipolaronic systems, exhibiting spatially symmetry broken states related to bipolaron 
ordering are:  Ti$_{4-x}$V$_x$O$_7$ \cite{LakkisPRB76} and Na$_x$V$_2$O$_5$ \cite{ChakravertyPRB78}. An example 
for a dilute bipolaron systems is WO$_{3-x}$ \cite{SchirmerJPC80}. In those systems 
the bipolarons form on adjacent cations, such as Ti$^{3+}$-Ti$^{3+}$, V$^{4+}$-V$^{4+}$ 
and W$^{5+}$-W$^{5+}$ bonds in strongly deformed octahedral ligand 
environments. Those bipolaronic units are imbeded in a corresponding 
background of  Ti$^{4+}$-Ti$^{4+}$, V$^{5+}$-V$^{5+}$ and  W$^{6+}$-W$^{6+}$ 
molecular units, which together with the bipolaronic entities, constitute those 
crystalline materials.

\subsection{Bipolaron dissociation and recombination in WO$_{3-x}$}

\begin{figure}
\begin{minipage}[b]{6.5cm}

\includegraphics[width=6.5cm]{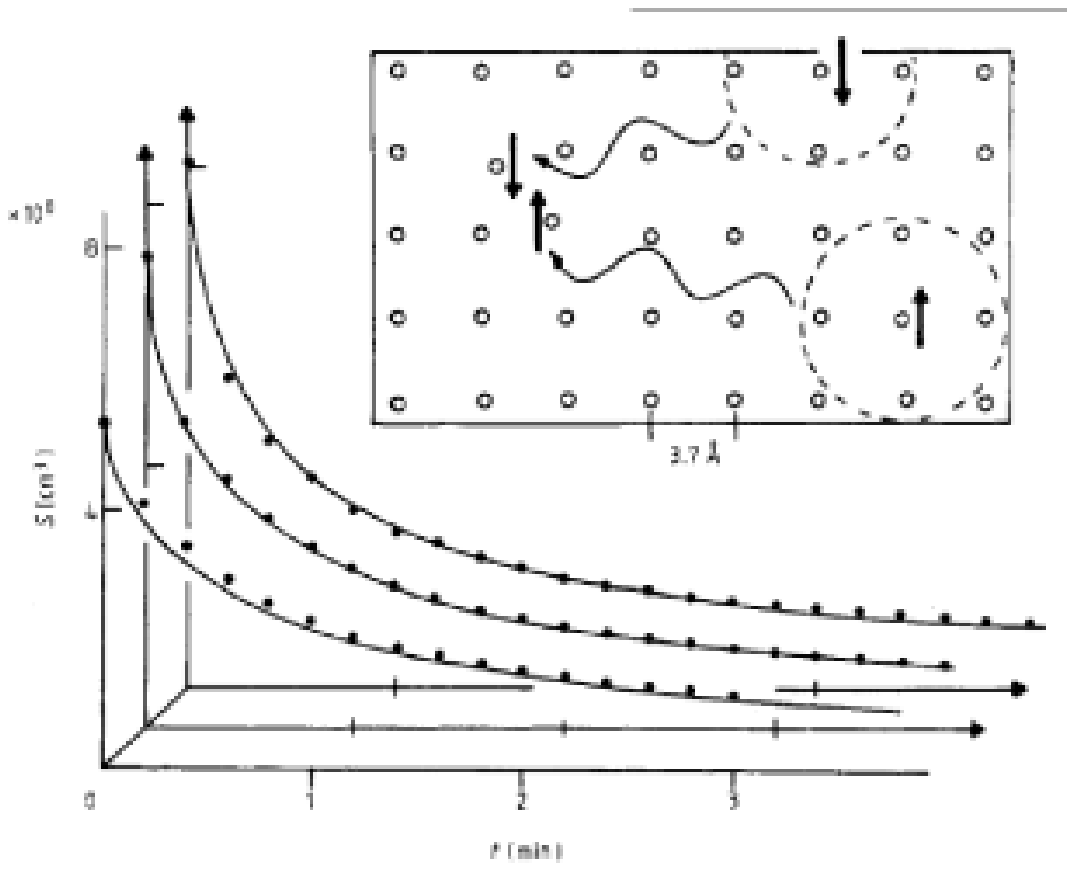}
\caption{Time evolution of the number of W$^{5+}$ sites (at different 
concentrations created with varying pump laser intensities breaking the  bipolaronic  
W$^{5+}$-W$^{5+}$ units) after after shutting off the pump-laser (after ref.~\cite{SchirmerJPC80}). The 
full lines represent the theoretical bimolecular relaxation behavior. The inset is a schematic picture for the bimolecular relaxation process.}
\label{Salje2}
\end{minipage}
\hspace{0.15cm}
\begin{minipage}[b]{6.5cm}
\includegraphics[width=6.5cm]{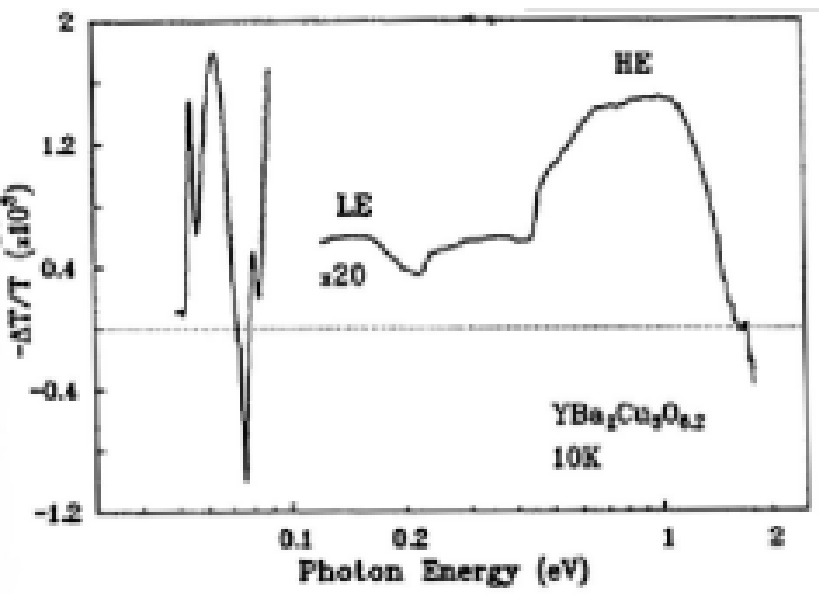}
\caption{The photoinduced infrared absorption coming from the laser-stimulated, 
respectively laser-suppressed  vibrational lattice modes 
(low frequency part) and the photomodulation response for electrical ingap 
excitations (high frequency part) in YBa$_2$Cu$_3$O$_{6.2}$ (after ref.~\cite{Taliani90}).}
\label{Taliani}
\end{minipage}
\end{figure}

Bipolarons can be dissociated with light which, in the case of  WO$_{3-x}$, 
leads to the creation of isolated W$^{5+}$ cation sites in vibrationally 
excited states of the ligand environments and whose concentration can be 
tracked by electron spin resonance (ESR) signals coming from those W$^{5+}$ sites. 
Optical absorption 
measurements \cite{SchirmerJPC80} in non-illuminated samples show a peak in the 
spectrum centered around 0.7~eV coming from pre-existing isolated  W$^{5+}$ cations 
sites. When the crystal is illuminated with a broad spectral band centered 
around a suitable frequency (1.1~eV in this case) the optical absorption coming 
from the W$^{5+}$ cations sites increases while, concomitantly, an intrinsic shoulder 
of the absorption band at around 1~eV---attributed to the absorption coming from the 
intrinsic bipolaronic W$^{5+}$-W$^{5+}$ bonds---decreases correspondingly. After the illumination is shut off, the single-polaron 
vibrationally exited W$^{5+}$ units relax to  
bipolaronic ones, as can be tracked by ESR measurements as a function of time (see fig.~\ref{Salje2}). 

The time evolution of this relaxation
follows a bimolecular recombination process controlled by a rate equation given by
\begin{equation}
{{\rm d}n_{\rm W}\over {\rm d}t} = -B(t) n^2_{\rm W}\, ,
\label{eq54}
\end{equation}
where $n_W$ denotes the density of the W$^{5+}$ ions and $B(t)$ is a 
function describing diffusion limited reactions. This relaxation behavior is 
clearly distinct from that of mono-molecular process which describe relaxation 
processes from vibrationally excited single polaron units to their ground state. Such 
time resolved studies permit to investigate the rebuilding of the phonon 
cloud for a bipolaron after its  separation into two separate polarons, much as would 
be expected for processes of bipolarons hopping between neighboring sites.

\subsection{Photo-induced polarionic charge carriers in high $T_{\rm c}$ cuprates}

A somewhat related in spirit technique was used in the study of polaronic 
features in the high $T_{\rm c}$ cuprates \cite{Taliani90}. It was based on examining 
the variation in 
the optical absorption spectrum upon doping those materials in the semiconducting 
phase with electron-hole  pairs using  so-called photo-induced doping. The idea was to 
use laser illumination with photon energies bigger than the semiconductor gap 
in the  pump laser beam. In the specific case of the high $T_c$ cuprates this meant 
a photon  energy  of typically 2 to 2.5~eV and which created an estimated density of 
charge carriers 
of about $10^{19}$ per cm$^3$. The injected photoinduced charge carriers modify 
locally the lattice symmetry and, by doing so, lead to the activation of 
corresponding local phonon modes (the 434~cm$^{-1}$ out of plane mode and the 
500~cm$^{-1}$ axial [Cu-O$_4$] stretching mode). These modes can be made evident as the 
steady state response of the system to the photo illumination in the absorption 
spectrum tested by a probe infrared beam (see fig.~\ref{Taliani}). In such an 
experimental set up, the 590~cm$^{-1}$ mode with  the 
tetragonal symmetry (being associated with the undistorted lattice which characterize  
the non-illuminated materials) loose in intensity (bleaching effect) 
in the absorption spectrum after the samples have been  illuminated. Examining the response 
of the system at higher frequencies, covering the energy range of the 
semiconducting gap and higher, by photo-modulation techniques (which 
test the photo-induced charge carriers by their absorption of light in a narrow  
frequency window) visualizes the  excitations with a life time which is 
inversely proportional to this frequency. The corresponding absorption spectra  
show an activation of charge excitations over a  
broad background, covering the energy region of the semiconducting gap, and a 
bleaching for energies above that frequency (see fig.~\ref{Taliani}). These results 
imply a shift of spectral weight upon illumination from the states above the gap in the non-illuminated  
samples into in-gap states. The  intensity of the absorption of the activated phonon modes 
as well as of  the activated electronic in-gap excitations turns out to scale like 
the square root of the intensity of the laser pulse illuminating the sample. This 
suggests that (in analogy with studies on WO$_{3-x}$ discussed above) those  excitations 
relax via  bimolecular recombination processes. Since the studies on photo-induced  
charge carriers have features similar to  those in  low doped systems, obtained by 
chemical substitution,  it has been tempting to conclude that the charge  carriers 
in the low doped superconducting samples have resonant bipolaronic features of a 
finite lifetime rather than corresponding to  well defined stable bipolaronic 
entities. Such entities can nevertheless condense into a superconducting state  
and are controlled by phase rather than amplitude fluctuations as discussed in my lecture: ``{\it From Cooper pairs to resonating bipolarons}'' in this volume.

\subsection{Bipolaronic charge ordering in Ti$_4$O$_7$}

\begin{figure}
\begin{minipage}[c]{6.5cm}
\includegraphics[width=6.5cm]{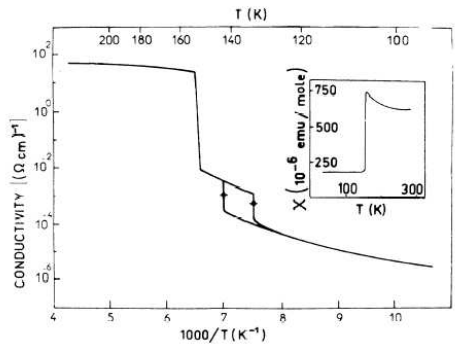}
\end{minipage}
\hspace*{0.15cm}
\begin{minipage}[c]{6.75cm}
\caption{The electrical conductivity for Ti$_4$O$_7$ (after ref.~\cite{LakkisPRB76}) showing 
with increasing $T$ a transition from a low temperature activated hopping regime to a different activated hopping regime at $T\simeq 140$~K followed by a second transition from that to a metallic phase at  $T \simeq 150$~K. The magnetic susceptibility, shown in the inset, does not change at the low temperature transition}
\label{Lakkis}
\end{minipage}
\end{figure}

A particularly interesting and physically very rich example of a dense bipolaronic system 
is  Ti$_4$O$_7$. It exhibits a low temperature phase ($T \leq 140$~K) where bipolarons 
are in a symmetry broken ordered state, consisting of diamagnetic 
$T^{3+}-T^{3+}$ bipolaronic diatomic pairs with sensibly reduced intra-molecular 
distances as compared to  $T^{4+} -T^{4+}$ pairs with which they alternate in  
quasi 2D slab like structures. Upon increasing the temperature, there is a small 
interval, [140~K~$\leq T \leq 150$~K], were these bipolaronic electron 
pairs are dynamically disordered and eventually break up into individual electrons 
and a metallic phase for $T \geq 150$~K. The experimental measurements of this material  
involved x-ray diffraction, resistivity, susceptibility and specific heat measurements 
as well as electron paramagnetic resonance (EPR) studies, summarized in ref.~\cite{LakkisPRB76}. The two 
low temperature phases in the regimes $T \leq 140$~K and 
140 K~$\leq T \leq 150$~K are semiconducting with a conductivity characterized by 
comparable activation energies of the order of 0.16~eV. The phase transition between 
those two semiconducting phases can be attributed to an order-disorder phase transition 
where the bipolarons are essentially the same as in the low temperature ordered 
phase, as evident from the absence of any change in  the magnetic 
susceptibility  as well as  intra-molecular distance of the bipolarons 
when going through this phase transition. The transition to the high temperature phase 
at $T$~= 150~K is characterized by a breaking up of the bipolarons  into itinerant 
electrons leading to a metal with an enhanced Pauli susceptibility and a substantial 
decrease in the unit-cell volume (see fig.~\ref{Lakkis}). 
Both transitions are first order and the 
entropy change in the high temperature transition is due, in roughly equal amounts,
to electronic and lattice  contributions. The most important feature 
of this system is however that the disordered bipolaronic phase is a dynamical 
rather than static disorder as evidenced from EPR experiments which show a vanishing of the  EPR line upon entering this phase from 
the low temperature ordered phase. It was on the basis of these experimental findings 
that the possibility of a condensation of bipolarons was initially 
proposed \cite{AlexandrovPRB81}. This broke with a traditionally severely guarded 
doctrine and stimulated to look for superconducting materials which (i) were oxides, (ii) have 
reduced dimensionality, (iii) are close to insulating parent compounds and (iv) are 
generally poor rather than good metals in the normal phase.

On the basis of such a scenario and the  Hostein model one could expect for such systems a superfluid state of bipolarons, albeit with a very small critical temperature, since being inversely proportional to the bipolaron mass which is typically several orders of magnitude bigger than the electron band mass. The real difficulty to observe this type of superfluidity in crystalline materials might however be related to the fact that the standard polaron models, generally based on harmonic lattice potentials, totally neglect any relaxation processes. High $T_c$ cuprates are clearly not candidates for this extreme case of Bipolaronic Superconductivity but are likely to contain localized bipolarons as resonant states inside the Fermi sea of itinerant electrons which could result in a superconducting  state controlled by phase rather than amplitude correlations.

\section{Summary}

 In this introductory lecture I briefly reviewed various kinds 
of electron-lattice couplings, characterizing different classes of materials and which give rise to two distinct categories of polarons: large Fr\"ohlich and small Holstein polarons. The importance of treating the phonons as quantum rather than classical variables became evident in connection with the question of itinerancy of the polaronic charge carriers. The dynamics of the 
polaron motion, exemplified in real time, shows a highly non-linear physics involving the dynamics of coupled charge and the lattice fluctuations which mutually drive each other. The present discussion was restricted to the single polaron respectively bipolaron problem and to limiting cases (such as strong coupling anti-adiabatic limit) where the Many Polaron problem can be decomposed into a band of single polaron states with different wave vectors. The fundamental questions of the polaron problem which pose themselves today, are evidently beyond the topics touched upon in this preliminary discussion. These are questions which concern the cross-over between the adiabatic and anti-adibatic regimes, the polaron induced residual interactions in a Many Polaron system and its 
dependence on the range of electron-lattice coupling as well as on the density of charge carriers, which can result in possible transitions between insulating and metallic behavior of polarons. At this stage, in order to tackle this kind  of problems we have to
resort to highly sophisticates numerical techniques, which will be presented in this school. The hope is that eventually this will give us some insight into this complex 
Many Body problem so that sooner or later we can formulate this polaron physics in a way where analytical approaches can capture its main qualitative features.

\end{document}